\documentclass[twocolumn,prl,showpacs,superscriptaddress]{revtex4}

\usepackage{amsmath,amssymb,amsfonts}
\usepackage{CJK}
\usepackage{bm}
\usepackage{array}
\usepackage{multirow}
\usepackage{graphicx}

\begin{document}

\title{Revealing a topological connection between stabilities of Fermi surfaces  and topological insulators/superconductors}


\author{Y. X. Zhao}
\email[]{zhaoyx@hku.hk}
\author{Z. D. Wang}
\email[]{zwang@hku.hk}

\affiliation{Department of Physics and Center of Theoretical and Computational Physics, The University of Hong Kong, Pokfulam Road, Hong Kong, China}


\date{\today}

\begin{abstract}
A topology-intrinsic connection between the stabilities of Fermi surfaces (FSs) and topological insulators/superconductors (TIs/TSCs) is revealed. In particular, a one-to-one relation between the topological types of FSs and TIs/TSCs is rigorously derived; combining it with a well-established topological theory of FSs, we produce a complete table illustrating precisely topological types of all TIs/TSCs, while a valid part of it was postulated before. Moreover, we  propose and prove a general index theorem that relates the topological charge of FSs on the natural boundary of a strong TI/TSC to its bulk topological number. Implications of the general index theorem on the boundary quasi-particles are also addressed. 
\end{abstract}
\maketitle

  For the past several years, the research on topological insulators (TIs) and superconductors (TSCs) has greatly attracted both theoretical and experimental interests, which becomes a hot spot of contemporary physics~\cite{Kane-RMP, XLQi-RMP}.
  As is known, various kinds of TIs/TSCs, which possess the time-reversal symmetry(TRS) or/and particle-hole symmetry(PHS), and chiral symmetry(CS), may be realised whenever the symmetry constraints on the bulk topology are enforced, while
 these symmetries can be preserved in the presence of weak disorders from a viewpoint of random matrix theory~\cite{Ramdom Matrix 0,Ramdom Matrix I,Ramdom Matrix II}. A TI/TSC, while it is fully gapped in the bulk spectrum, can have robust gapless modes on its boundary, being distinctly different from those in  an ordinary insulator or superconductor. According to a conventional wisdom, these gapless modes are originated from a fact that the boundary of a TI/TSC as a domain wall separates the system in different topological phases, with topological order-parameters being identified as the bulk topological numbers. 
  As it is generally believed that there exists a faithful bulk-boundary correspondence, the above wisdom is practically useful, while it is likely to provide a qualitative insight on topology merely from the bulk to boundary, rather than a complete and quantitative description of this topological  correspondence. 
  On the other hand,  stimulated by Volovik's pioneering work on topological Fermi surfaces
 (FSs) without any discrete symmetry~\cite{Volovik-Book}, a  topological theory of FSs has recently been established for describing many-fermion systems with TRS or/and PHS, where a nontrivial topological charge of an FS with a specific symmetry ensures its topological stability against symmetry-preserving weak disorders/perturbations and interactions~\cite{FS-Classification}. 
 Therefore, it is natural and insightful to look into directly the topological properties of boundary FSs of a TI/TSC and to reveal quantitatively their intrinsic connection to the bulk topology, particularly considering that the robustness of these gapless boundary modes may have the same topological origin as that of the bulk.
This new sort of boundary-to-bulk insight on the topological essence of TIs/TSCs may further deepen our understanding of the bulk-boundary correspondence, supplementing the conventional one in
 a more efficient way because only information in the vicinity of FSs is actually needed to disclose the topological character. 
      
   In this Letter, based on a well-established topological theory of FSs~\cite{FS-Classification}, we first derive a one-to-one relation between the topological types of FSs and TIs/TSCs, from which we establish a complete table (Tab.[\ref{tab:Periodic-Table}] in a right-to-left manner) that illustrates precisely topological types of all TIs/TSCs {\it {vs}} the bulk dimension and symmetry class. Intrigued by this relation, we propose a general index theorem [Eq.(\ref{eq:Index-theorem})] as a quantitive description for the topological boundary-bulk correspondence of TIs/TSCs and prove it in the framework of Dirac-matrix construction. Finally, we address briefly the predictions and restrictions on the boundary low-energy effective theories of TIs/TSCs.

 Let us begin with a brief introduction to topological FSs. An FS is actually a region of fermionic gapless modes in the $\mathbf{k}$-space of energy spectrum. Some FSs are stable against disorders/perturbations, while some others are vulnerable and easy to be gapped~\cite{Volovik-Book}. It is found that the topological charge of an FS is responsible for its stability~\cite{Volovik-Book, Horava, Volovik-Vacuum,FS-Classification}. For a $(d_s-p)$-dimensional FS in a $d_s$-dimensional $\mathbf{k}$-space, we can choose a $p$-dimensional sphere $S^p$ from $(\omega,\mathbf{k})$-space to enclose it from its transverse dimensions, where $p$ is referred to as the codimension of the FS. Note that the spectrum is gapped on the whole $S^p$ since the $S^p$ is constructed in the transverse dimensions of an FS. For an FS without any discrete symmetry, 
its topological charge is given by the homotopy number of the inverse Green's function restricted on the $S^p$~\cite{Volovik-Book, Horava, Volovik-Vacuum}, i.e., $G^{-1}(\omega,\mathbf{k})|_{S^p}=[i\omega-\mathcal{H}(\mathbf{k})]|_{S^p}$. This idea has recently been generalized to FSs of the other classes for characterising the corresponding types of symmetry-dependent topological charges~\cite{FS-Classification}, as summarized in Tab.[\ref{tab:Periodic-Table}] (in a left-to-right manner), where the $\mathbf{Z}$, $\mathbf{Z}^{(1)}_2$, and $\mathbf{Z}^{(2)}_2$ denote the integer-valued topological charge,  $\mathbf{Z}_2$-valued (integers of modular $2$, i.e., $0$ or $1$ ) topological charge for the first descendant of a $\mathbf{Z}$-type, and  $\mathbf{Z}_2$-valued topological charge for the second descendant of a $\mathbf{Z}$-type.

\begin{table}
\begin{centering}
\begin{tabular}{|c|c|c|c|c|c|c|c|c|c|}
\hline 
FS & AI & BDI & D & DIII & AII & CII & C & CI & $\frac{\mathrm{TI}}{\mathrm{TCS}}$\tabularnewline
\hline 
\hline 
T & +1 & +1 & 0 & -1 & -1 & -1 & 0 & +1 & T\tabularnewline
\hline 
C & 0 & +1 & +1 & +1 & 0 & -1 & -1 & -1 & C\tabularnewline
\hline 
S & 0 & 1 & 0 & 1 & 0 & 1 & 0 & 1 & S\tabularnewline
\hline 
\hline 
p\textbackslash{}i & 1 & 2 & 3 & 4 & 5 & 6 & 7 & 8 & i/d\tabularnewline
\hline 
1 & 0 & 0 & $\mathbf{Z}$ & $\mathbf{Z}_{2}^{(1)}$ & $\mathbf{Z}_{2}^{(2)}$ & 0 & $2\mathbf{Z}$ & 0 & 2\tabularnewline
\hline 
2 & 0 & 0 & 0 & $\mathbf{Z}$ & $\mathbf{Z}_{2}^{(1)}$ & $\mathbf{Z}_{2}^{(2)}$ & 0 & $2\mathbf{Z}$ & 3\tabularnewline
\hline 
3 & $2\mathbf{Z}$ & 0 & 0 & 0 & $\mathbf{Z}$ & $\mathbf{Z}_{2}^{(1)}$ & $\mathbf{Z}_{2}^{(2)}$ & 0 & 4\tabularnewline
\hline 
4 & 0 & $2\mathbf{Z}$ & 0 & 0 & 0 & $\mathbf{Z}$ & $\mathbf{Z}_{2}^{(1)}$ & $\mathbf{Z}_{2}^{(2)}$ & 5\tabularnewline
\hline 
5 & $\mathbf{Z}_{2}^{(2)}$ & 0 & $2\mathbf{Z}$ & 0 & 0 & 0 & $\mathbf{Z}$ & $\mathbf{Z}_{2}^{(1)}$ & 6\tabularnewline
\hline 
6 & $\mathbf{Z}_{2}^{(1)}$ & $\mathbf{Z}_{2}^{(2)}$ & 0 & $2\mathbf{Z}$ & 0 & 0 & 0 & $\mathbf{Z}$ & 7\tabularnewline
\hline 
7 & $\mathbf{Z}$ & $\mathbf{Z}_{2}^{(1)}$ & $\mathbf{Z}_{2}^{(2)}$ & 0 & $2\mathbf{Z}$ & 0 & 0 & 0 & 8\tabularnewline
\hline 
8 & 0 & $\mathbf{Z}$ & $\mathbf{Z}_{2}^{(1)}$ & $\mathbf{Z}_{2}^{(2)}$ & 0 & $2\mathbf{Z}$ & 0 & 0 & 9\tabularnewline
\hline 
$\vdots$ & $\vdots$ & $\vdots$ & $\vdots$ & $\vdots$ & $\vdots$ & $\vdots$ & $\vdots$ & $\vdots$ & $\vdots$\tabularnewline
\hline 
\end{tabular}
\par\end{centering}

\caption{Cartan Classification of systems, and topological classification of  FSs and TIs/TSCs~\cite{FS-Classification}.\label{tab:Periodic-Table} In the upper part of the table, T, C, and S denote TRS, PHS, and CS, respectively. 0 indicates the absence of the corresponding symmetry, $\pm$1 indicates the sign of TRS or PHS, and 1 indicates the existence of CS. In the lower part,  $i$ is the index of symmetry classes, $p$ is the codimension of an FS, and $d$ is the spatial dimension of a TI/TSC. The elements $\mathbf{Z}$, $2 \mathbf{Z}$, $\mathbf{Z}^{(1)}_2$, and $\mathbf{Z}^{(2)}_2$ in the above periodic (eight-fold) table represent the corresponding topological types, respectively.}
\end{table}


We now derive 
the relation between the classification of FSs and that of TIs/TSCs, as indicated in Tab.[\ref{tab:Periodic-Table}]. 
First, It is crucial to observe that all six formulas for the topological charges of FSs  can formally be used to calculate the topological numbers of TIs/TSCs, where we make the integration over the whole $(\omega,\mathbf{k})$-space in these formulas\cite{XLQi-RMP, Topological-Number-I,Topological-Number-II}, in stead of making it over $S^p$ for an FS, as illustrated in Supplemental Material~\cite{Supp}.
\begin{figure}
\includegraphics[scale=0.25]{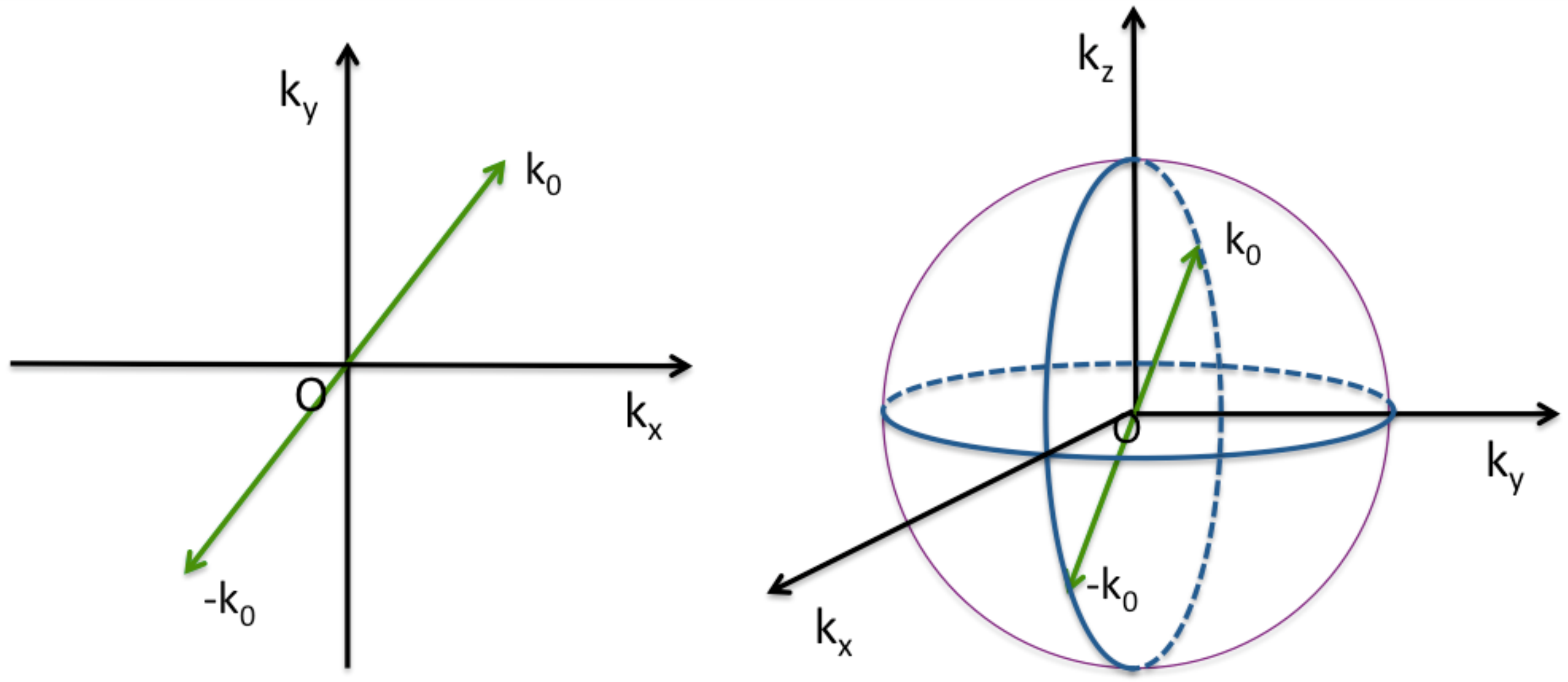}
\caption{Symmetry Identification. The left corresponds to a two-dimensional $\mathbf{k}$-space, and the right corresponds to an $S^2$ enclosing a Fermi point in a three-dimensional $\mathbf{k}$-space. \label{fig:Symmetry-Identification}}
\end{figure}
We emphasise here that the operation of TRS and PHS on the $(\omega,\mathbf{k})$-space of a TI/TSC is essentially different from that on an $S^p$ for an FS,  as shown in Fig.[\ref{fig:Symmetry-Identification}]. If the symmetry operation is regarded as an identification, the resulting topological space of $S^p$ for an FS is a $p$-dimensional projective space $RP(p)$~\cite{rp}, while that of $\mathbf{k}$ space is not. This topological difference of symmetry operation leads to that the topological types of TIs/TSCs {\it{vs.}} the dimension $d$ have globally a one-dimension shift with respect to those of FSs {\it{vs.}} the codimension $p$, $i.e.$, 
\begin{equation}
K_{TI}(d,i)=K_{FS}(d-1,i),\label{eq:Table-relation}
\end{equation}
where $K_{FS}(p,i)$ and $K_{TI}(d,i)$ denote, respectively, the topological types of FSs (with the codimension $p$) and TIs/TSCs (with the spatial dimension $d$) in the $i$th class.   Without loss of generality, we below consider class AII ($i=5$) and highlight the key steps in the derivation of this relation. It is known that in the absence of any discrete symmetry, FSs with nontrivial topological charges can exist only when $p=2n-1$ with $n$ being an integer~\cite{Volovik-Book, Horava}, and TIs/TSCs can have nontrivial topological numbers only when $d=2n$. After operations  of the minus-sign TRS  on $S^p$ and $(\omega,\mathbf{k})$ space~\cite{Supp}, we obtain the following relations for FSs and TI/TSCs:
\[
\begin{cases}
\nu_Z (2n+1,5)=(-1)^{n-1}\nu_Z (2n+1,5)\\
N_Z (2n,5)=(-1)^{n}N_Z (2n,5)
\end{cases},
\]
where $\nu_Z(p,i)$ and $N_Z(d,i)$ denote respectively the topological charge of an FS (with the codimension $p$) and the topological number of a TI/TSC (with the dimension $d$) in the $i$th class if they are of $\mathbf{Z}$-type. It is found for class AII that  nontrivial $\mathbf{Z}$-type (or $2 \mathbf{Z}$-type) FSs can exist only when $p=4m-1$, while there are nontrivial $\mathbf{Z}$-type (or $2 \mathbf{Z}$-type) TIs/TSCs only if $d=4m$,  where $m$ is an integer. Since $\mathbf{Z}^{(1)}_2$ and $\mathbf{Z}^{(2)}_2$, as the first and second descendants of $\mathbf{Z}$, will follow $\mathbf{Z}$ successively when $d$ is reduced, our remaining task is to distinguish types of $\mathbf{Z}$ and $2\mathbf{Z}$ for TIs/TSCs, which can be done for class AII as follows: noticing that  FSs are of $\mathbf{Z}$-type when $p=8(m-1)+3$,
while TIs/TSCs, whose Hamiltonians can be constructed by Dirac matrices, possess unit topological number (i.e., $\mathbf{Z}$-type) for $d=8(m-1)+4$~\cite{Topological-Number-I, Supp}. Thus we have $K_{TI}(d,5)=K_{FS}(d-1,5)$. Since similar derivations can be made for the other seven classes (not shown here), Eq.(\ref{eq:Table-relation}) is rigorously deduced
and therefore the classification table of TIs/TSCs is established, as presented in Tab.[\ref{tab:Periodic-Table}] in a right-to-left manner, which is one of our main results. 



Notably, a similar classification table of TIs/TSCs was first postulated by generalizing a Pruiskein's theory for the integer quantum hall effect(IQHE)(without any above-mentioned symmetry) to the other symmetry classes~\cite{IQHE,TIClassification-I, TIClassification-II, TIClassification-III}. The Pruiskein's theory is a non-linear $\sigma$ model handling disorders, which attributes the stability of the gapless boundary modes to a topological Wess-Zumino-Witten term on the boundary~\cite{IQHE, WZ-term}. As a specific symmetry class has a unique group for its non-linear sigma model~\cite{Ramdom Matrix 0,Ramdom Matrix I,Ramdom Matrix II}, the classification in Tab[\ref{tab:Periodic-Table}] was postulated according to whether eligible topological terms can appear on the boundary of a system in each case~\cite{Topo-Terms}. While in the present work, first, it is shown rigorously that all types of the bulk topological numbers are just those (with one-to-one correspondence) for topological charges of FSs identified in Ref.~\cite{FS-Classification}, as seen in Eq.(\ref{eq:Table-relation}); secondly, the periodic table for TIs/TSCs is refined by classifying the two different topological types of $\mathbf{Z}^{(1,2)}_2$, enabling us to choose a corresponding  formula for calculating the topological charges correctly; while no such distinction can be seen from the existing results~\cite{TIClassification-I, TIClassification-II, TIClassification-III, K-theory}.

It has been known from many examples that the boundary gapless modes of a given TI/TSC with one (or more) above symmetry are robust against disorders/perturbations that do not break the corresponding symmetry. This implies that the FSs corresponding to these gapless modes are actually protected by their nontrivial topological charge with the same symmetry class according to a theory of topologically stable FSs~\cite{FS-Classification}, while such robustness can also be 
attributed to a nontrivial topological number of bulk. Therefore, we attempt to find a quantitative relation between the topological number $N$ of a bulk TI/TSC and the topological charge $\nu$ of its boundary FSs.
Let us consider a $d$-dimensional semi-infinite TI/TSC of a symmetry class $i$ with a boundary being at $x=0$ on its left-side.
Under the natural boundary condition of the TI/TSC, i.e., no dramatic anisotropy is induced and the symmetries in the $i$th class are preserved 
when the system approaches to its boundary, FSs on the boundary will always be some Fermi points in a topological sense.
This conclusion is obviously valid for $d=1, 2$. In addition, its validity can also be extended to TIs/TSCs for $d>2$~\cite{note}.
As a result,  the boundary codimension $p=d-1$ .
Then by appropriately choosing an $S^{d-1}$ in the $(\omega,\mathbf{k})$ space of the boundary to enclose all the FSs, we are able to compute the total topological charge $\nu(d-1,i)$ of all FSs. Therefore, from Eq.(\ref{eq:Table-relation}), which enables the same type of nontrivial topological charge of boundary FSs and bulk number to protect the FSs, 
we may expect a general index theorem as a quantitive topological description of the boundary-bulk correspondence of a TI/TSC:
\begin{equation}
\nu(d-1,i)=N(d,i). \label{eq:Index-theorem}
\end{equation}
Clearly, the above equation reflects essentially that the same topological origin protects both the stability of FSs on the boundary and the gapped spectrum of bulk against disorders/perturbations. We note that Eq.(\ref{eq:Index-theorem}) accounts for the earlier statements for classification of TIs/TSCs based on the random-matrix theory by assuming certain boundary topological terms in a non-linear $\sigma$ model~\cite{TIClassification-I,TIClassification-II,TIClassification-III}, and reveals that  
the bulk topology can  also be quantified by the topological charge of boundary FSs. 
Notably, a similar index theorem for systems without any symmetry was given in Ref.\cite{Volovik-Book}, and has been generalized to other cases with either physical arguments or strict verifications for some concrete models~\cite{Index-I,Index-II, Dirac-Models}, though the topological charge of FSs was not used explicitly. 

To see 
Eq.(\ref{eq:Index-theorem}) more clearly, 
we first present two typical examples, with the proof to be given later. Consider a 3-dimensional TI described by
the Dirac-lattice model 
\cite{Topological-Number-I, Supp}, 
$i.e.$ $\mathcal{H}_D=\sum^3_{i=1}\sin k_i\Gamma^{i+2}+(3+\sum_{i=1}^3\cos k_i)\Gamma^1$ with $\Gamma_i(i=1,\cdots,5)$ being $4\times 4$ hermitian Dirac matrices\cite{Dirac-Matrix, Supp} and $T=\Gamma^1\Gamma^3$. This model has TRS with a minus sign and thus  may have a non-trivial 
$\mathbf{Z}_2^{(1)}$-type, as seen from  Tab.[\ref{tab:Periodic-Table}]  $(d=3)$. On the other hand,
its boundary effective model reads $\mathcal{H}=k_z\sigma_2-k_y\sigma_3$ with $\sigma_i$ being Pauli matrices~\cite{Dirac-Models}, which preserves the same TRS and
has a Fermi point with codimension $p=2$.  
One can find that $\nu(2,5)=1=N(3,5)$~\cite{Supp}, verifying Eq.(\ref{eq:Index-theorem}). Similarly, another 2-dimensional model, which describes the quantum spin Hall effect, 
is given by $\mathcal{H}_{spin}=\sum^2_{i=1}\sin k_i\Gamma^{i+3}+(3+\sum_{i=1}^2\cos k_i)\Gamma^1$. It has also TRS with a minus sign and therefore belongs to class AII with a $\mathbf{Z}_2^{(2)}$-type topological number. Its boundary effective Hamiltonian is $\mathcal{H}=\sigma_zk_z$, which has a Fermi point ($p=1$) with the same $\mathbf{Z}_2^{(2)}$-type topological charge. We can also find that $\nu(1,5)=1=N(2,5)$~\cite{Supp}. 



At this stage, we turn to brief readers the proof of the general index theorem: Eq.(\ref{eq:Index-theorem}), with more details being presented in Ref.\cite{Supp}. 
First,  it is  seen from the  topological nature of a TI/TSC that the total topological charge of FSs on its boundary and the topological number of its bulk are both invariant under continuous deformations of its Hamiltonian without closing the bulk spectrum gap.  
Secondly, it is noted that Hamiltonian of each type of FS (point) for a given codimension $p$ and class $i$ with a unit charge (or double-unit charge for $2\mathbf{Z}$ type) can be expressed by Dirac matrices in a unified form:
\begin{equation}
\mathcal{H}_{FS}(\mathbf{k})=\sum_{a=1}^{p}k_{a}\Gamma_{(2n+1)}^{a+b},\label{Typical-FS}
\end{equation}
where 
$\Gamma^{a}_{(2n+1)}$ are $2^n\times2^n$ Dirac matrices, and  $b$ is an integer that is specified by the corresponding topological type of the FS. Thirdly,  a modified Dirac-type model for each type of bulk TI/TSC
 with unit topological number(or double-unit for $2\mathbf{Z}$ type) can also be constructed in a unified form:
\begin{equation}
\mathcal{H}_{TI}(\mathbf{k})=\sum_{a=1}^{d}k_{a}\Gamma^{a+b-1}_{(2n+1)}+\left(m-\epsilon k^{2}\right)\Gamma^{\alpha}_{(2n+1)},\label{Typical-TI}
\end{equation}
where $\epsilon$ is a constant,  $\Gamma^\alpha=i\Gamma^1\Gamma^2\Gamma^3$ for $2\mathbf{Z}$ cases and $\Gamma^\alpha=\Gamma^1$ otherwise. Note that $\epsilon$ term is necessary for a prescription of the singularity of $\mathcal{H}_{TI}$ when $k$ approaches to infinity~\cite{Volovik-Book}. This model has a unified expression of the topological number for all types of TIs/TSCs:
\begin{equation}
N(d,i)=(1\,\,\,or\,\,\,1/2)(\mathbf{sgn}(m)+\mathbf{sgn}(\epsilon)),
\end{equation}
where the coefficient is $1$ for $2\mathbf{Z}$ cases, and $1/2$ otherwise. 
Next, we proceed to consider an arbitrarily given TI/TSC in a certain type with the topological number $N(d,i)$, which can be deformed continuously to a model that is $N$-multiple (or for $2\mathbf{Z}$ cases, $(N/2)$-multiple) of the model (\ref{Typical-TI}), preserving the corresponding symmetries and without closing the bulk spectrum gap. After this deformation, both the bulk topological number and the total topological charge of boundary FSs are preserved, as indicated above~\cite{Note-Deformation}, and therefore it is sufficient to consider the unit model (\ref{Typical-TI}) for proof of Eq.(\ref{eq:Index-theorem}). The advantage of this deformation lies in that the boundary low-energy effective theory of the model (\ref{Typical-TI}) can be systematically derived by using a standard perturbation theory of quantum mechanics under the open boundary condition\cite{Supp}. After some tedious derivations detailed in Ref.~\cite{Supp}, the boundary effective theory of the model (\ref{Typical-TI}) turns out to be
\begin{equation}
\mathcal{H}_{eff}=\frac{1}{2}\left(\mathbf{sgn}(m)+\mathbf{sgn}(\epsilon)\right)\sum_{a=2}^{d}k_{a}\Gamma_{(2n-1)}^{a+b-3},\label{BET}
\end{equation}
for the boundary at $x=0$. Note that the symmetries of the system are now represented in the boundary effective theory (\ref{BET}). By matching the symmetry representation of the theory (\ref{BET}) to that of our constructed model (\ref{Typical-FS}), we find that not only the topological charge of Fermi point in the boundary effective theory (\ref{BET}) is equal to the topological number of the model (\ref{Typical-TI}), but also the topological type of the FS is the same as that of the bulk model (\ref{Typical-TI}), namely, the general index theorem is validated\cite{Supp}. 

Before concluding this paper, we address briefly implications of Eq.(\ref{eq:Index-theorem}) on the form of a boundary effective theory with a given bulk topological number. Based on the Atiyah-Bott-Singer(ABS) construction in K-theory and a fact that an FS of multiple charge can be perturbed to a more stable state consisting of a number of unit FSs\cite{K-theory-II,K-theory-III, Horava}, we assert that a typical low-energy  effective theory of boundary for a $\mathbf{Z}$ type TI/TSC is a collection of $\sum_{i=1}^{2n+1}k_{i}\Gamma_{(2n+1)}^{i}$ for non-chiral cases or $\sum_{i=1}^{2n}k_{i}\Gamma_{(2n+1)}^{i}$ for chiral cases, with the total topological charge being equal to the bulk topological number. On the other hand, for $\mathbf{Z}^{(1,2)}_2$ type TIs/TSCs, the constraint from ABS construction is not as strong as that of the above cases, while the constraint  on $\mathbf{Z}^{(1)}_2$ type is stronger than that on $\mathbf{Z}^{(2)}_{2}$ type, since the latter needs one more extra parameter. Nevertheless, a boundary effective theory can still be constructed by Dirac matrices, whose expression is not uniquely determined.
 For instance, $\mathcal{H}=k_z\sigma_3+\alpha k^{(2n+1)}_z+\beta k^{(2m+1)}_z$ and $\mathcal{H}=k^3_z\sigma_3+\alpha k^{(2n+1)}_z+\beta k^{(2m+1)}_z$ are both eligible as the boundary effective theory of a quantum spin hall system.
 
 In summary, the topology-intrinsic connection between the stabilities of FSs and TIs/TSCs has been revealed.  
 Furthermore,
 we have proposed and proven a general index theorem, providing for the first time a quantitive description of the bulk-boundary correspondence of TIs/TCSs. 
 It is anticipated that the present work will also provide a deeper insight and open a wider door for exploring exotic boundary gapless modes of TIs/TSCs such as Majorana fermions.

\begin{acknowledgments}
We thank A. J. Leggett and G. E. Volovik for helpful discussions. This work was supported
by the GRF (HKU7058/11P\&HKU7045/13P),  the CRF (HKU8/11G) of Hong Kong, and the URC fund of HKU.
\end{acknowledgments}

\section{Supplemental Material}

\section{SI. Derivation of Eq.(1)}
In the derivation, the first key point is that for a $\mathcal{H}(\mathbf{k})$ with either TRS or PHS, the symmetries can be preserved on the $S^p$ if it is chosen to be centrosymmetric with respect to the origin of $(\omega,\mathbf{k})$-space. Secondly, It is crucial to observe that all six formulas for the topological charges of FSs  can formally be used to calculate the topological numbers of TIs/TSCs, where we make the integration over the whole $(\omega,\mathbf{k})$-space, in stead of making it over $S^p$ for an FS. 
The topological difference of symmetry identification plays an essential role in the derivation of the topological classification table of TIs/TSCs. In practical calculations for FSs, we use the following spherical coordinates for the
chosen $S^{p-1}$,
\begin{equation}
\begin{cases}
k_1=\cos s_{1}\\
k_2=\sin s_{1}\cos s_{2}\\
k_3=\sin s_{1}\sin s_{2}\cos s_{3}\\
\,\,\,\,\,\,\,\,\,\,\,\, \vdots\\
k_{p-1}=\sin s_{1}\sin s_{2}\cdots\sin s_{p-2}\cos s_{p-1}\\
k_p=\sin s_{1}\sin s_{2}\cdots\sin s_{p-2}\sin s_{p-1}
\end{cases}\label{Spherical-Coordinates}
\end{equation}
with $s_{i}\in[0,\pi]$ for $i=1,2,\cdots,p-2$ and $s_{p-1}\in[0,2\pi)$. Without loss of generality, 
we have assumed that the FS is a Fermi point with codimension $p$ in a $(p+1)$-dimensional $(\omega,\mathbf{k})$-space.
In the following, we focus on the class AII as an illustration, which has only TRS with
$\eta_{T}=-1$, while other seven classes can be treated in the same way and  results are the same. The TRS on the chosen $S^{p-1}$ with the spherical coordinates can be represented as 
\[
T^{\dagger}\mathcal{H}(s)T=\mathcal{H}^{T}(\pi-s_{1},\cdots,\pi-s_{p-2},\pi+s_{p-1}).
\]
Note that the transformation of $s_{p-1}$ is different from those of the others.
Correspondingly, Green's function $G(\omega,\mathbf{k})=\left[i\omega-\mathcal{H}(\mathbf{k})\right]^{-1}$ is
transformed as
\begin{equation}
T^{\dagger}G(\omega,s)T=G^{T}(\omega,\pi-s_{1},\cdots,\pi-s_{p-2},\pi+s_{p-1}).\label{eq:TRS-FS0}
\end{equation}
While  the Green's function for a TI in the class AII satisfies
\begin{equation}
T^{\dagger}G(\omega,\mathbf{k})T=G^{T}(\omega,-\mathbf{k}).\label{eq:TRS-TI}
\end{equation}
It is  seen that that all $\mathbf{k})$s 
of the TI reverse the signs under the TRS transformation, which is in contrast to the situation of the chosen $S^{p-1}$ enclosing
an FS where the last coordinate $s_{p-1}$ does not reverse its sign. 
We first illustrate how
TRS makes the $\mathbf{Z}$-type topological charge of an FS with codimension $p=4m+1$ vanishing. The formula
for the $\mathbf{Z}$-type topological charge with codimension $p=2n+1$ is given by
\begin{eqnarray}
 &  & \nu(2n+1,5)\nonumber \\
 & = & C_{2n+1}\int_{S^{p}} d\omega d^{2n}s\;\epsilon^{\mu_{1}\mu_{2}\cdots\mu_{2n+1}}\nonumber\\
 &  & \mathbf{tr}\left(G\partial_{\mu_{1}}G^{-1}G\partial_{\mu_{2}}G^{-1}\cdots G\partial_{\mu_{2n+1}}G^{-1}(\omega,s)\right),\label{eq:Topo-Charge}
\end{eqnarray}
where $C_{2n+1}=-n!/(2n+1)!(2\pi i)^{n+1}$. Implementing the TRS transformation, $i.e.$ Eq.(\ref{eq:TRS-FS0}),
and the coordinate substitution $s'_i=\pi-s_i$ with $i=1,2,\cdots,2n-1$ and $s'_{2n}=\pi+s_{2n}$, we have
\begin{eqnarray}
 &  & \nu(2n+1,5)\nonumber\\
 & = & -C_{2n+1}\int_{S^{p}} d\omega d^{2n}s'\;\epsilon^{\mu'_{1}\mu'_{2}\cdots\mu'_{2n+1}}\nonumber\\
 &  & \mathbf{tr}\left(G^{T}\partial_{\mu'_{1}}G^{T-1}\cdots G^{T}\partial_{\mu'_{2n+1}}G^{T-1}(\omega,s')\right)\nonumber\\ 
 & = & C_{2n+1}\int_{S^{p}} d\omega d^{2n}s\;\epsilon^{\mu_{1}\mu_{2}\cdots\mu_{2n+1}}\nonumber\\
 &  & \mathbf{tr}\left(G\partial_{\mu_{2n+1}}G^{-1}G\partial_{\mu_{2n}}G^{-1}\cdots G\partial_{\mu_{1}}G^{-1}(\omega,s)\right)\label{eq:Partial-sign}.
\end{eqnarray}
Making a permutation that reverses the order of all the indices of $\epsilon$, we obtain 
\begin{equation}
\nu(2n+1,5)=(-1)^{n-1}\nu(2n+1,5),\label{eq:TRS-FS}
\end{equation}
where the extra $\left(-1\right)^{n}$ comes from the permutation. Now it is clear that when $n=2m$ with $m$ being an integer, $i.e.$ $p=4m+1$,
the $\mathbf{Z}$-type topological charge of an FS is vanished for the class
AII. 
For the corresponding case of TI, the procedure is almost the same. First, the integral form of topological number is the same as Eq.(\ref{eq:Topo-Charge}), but the integration is made over the whole $(\omega,\mathbf{k})$-space: 
\begin{eqnarray}
 &  & N(2n,5)\nonumber \\
 & = & C_{2n+1}\int d\omega d^{2n}k\;\epsilon^{\mu_{1}\mu_{2}\cdots\mu_{2n+1}}\nonumber\\
 &  & \mathbf{tr}\left(G\partial_{\mu_{1}}G^{-1}G\partial_{\mu_{2}}G^{-1}\cdots G\partial_{\mu_{2n+1}}G^{-1}(\omega,\mathbf{k})\right). \label{eq:Topo-Number}
\end{eqnarray}
In a similar way to the case of FS, 
we make the variable substitution $\mathbf{k'}=-\mathbf{k}$, matrix transposition, and then a permutation to reverse all the indices of $\epsilon$. The only difference is that when substituting the variables, we could not obtain an extra minus sign as that in the second equality of Eq.(\ref{eq:Partial-sign}), since the number of partial derivatives of $k_i$ is even. Thus we have
\begin{equation}
N(2n,5)=\left(-1\right)^{n}N(2n,5), \label{eq:TRS-TI}
\end{equation}
which implies the topological number of a TI with $d=4m+2$ in the class AII is always trivial, but that with $d=4m$
can be nontrivial. 
      
Comparing Eq.(\ref{eq:TRS-TI}) with Eq.(\ref{eq:TRS-FS}), we see clearly a one-dimension shift from $K_{FS}(p, 5)$ to $K_{TI}(d,5)$. From the established classification table of FSs in class AII, we know that a sole but minor remaining uncertainty is whether the topological types of TIs/TSCs for 
$d=8m+4$ are $\mathbf{Z}$ or $2\mathbf{Z}$, which can surely be fixed by examining an appropriately chosen case. Since the Dirac model of $d=4$ can have a unit topological charge belonging to $\mathbf{Z}$-type,  the topological charges for $d=8m+4$ is of $\mathbf{Z}$-type, and thus $K_{FS}(d-1,5)=K_{TI}(d,5)$. The same results are obtained for the other seven classes due to the same reason. Thus we show rigorously that $K_{FS}(d-1,i)=K_{TI}(d,i)$, which reflects the topological difference between the $\mathbf{k}$-space and the chosen $S^{p-1}$ to enclose an FS when the symmetry class is concerned.

\section{SII. Topological charge of Fermi points on the boundary of TIs}

\subsection{SII.A. Topological charge of the Fermi point on the boundary of 3-dimensional TI}

The boundary effective theory for a 3-dimensional TI reads
\[
\mathcal{H}(\mathbf{k})=k_{y}\sigma_{1}-k_{x}\sigma_{2}+\alpha (k_{x}+k_{y})\sigma_{3},
\]
where the term $\alpha(k_{x}+k_{y})\sigma_{3}$ merely indicates that the Hamiltonian in the class AII allows such term, which will turn out to be irrelevant for the topological charge of the Fermi point. This Hamiltonian has a TRS with sign $-$, $i.e.$~\cite{FS-Classification},
\[
\mathcal{H}(\mathbf{k})=\sigma_{2}\mathcal{H}^{T}(-\mathbf{k})\sigma_{2}^{-1},\quad\sigma_{2}^{T}=-\sigma_{2}.
\]
It belongs to the class AII. The Fermi
surface is the point $\mathbf{k}=0$ with the codimension
$2$. According to Ref.\cite{FS-Classification}, we can calculate its $\mathbf{Z_{2}^{(1)}}$-topological charge from Eq.(6) there. In order to do this, we choose an $S^{2}$ in the $(\omega,\mathbf{k})$-space enclosing the Fermi point, and make the following continuous extension of the Green's function restricted on the $S^{2}$:
\begin{eqnarray*}
G^{-1} & = & i\omega-\Delta\sigma_{3}\sin\theta\\
 &  & -\left[k\sin\phi\sigma_{1}-k\cos\phi\sigma_{2}+k(\cos\phi+\sin\phi)\sigma_{3}\right]\cos\theta ,
\end{eqnarray*}
where  $\phi\in[0,2\pi]$ parametrizes the circle in the $\mathbf{k}$-space enclosing the Fermi surface,  $\theta\in[0,\pi/2]$ is the parameter for extension, and $\Delta$ is a positive
constant. Substituting this Green's function into Eq.(6) in Ref.\cite{FS-Classification}, we obtain
\begin{eqnarray*}
\nu(2,5) & = & \frac{1}{12\pi^{2}}\int d\omega d\phi d\theta\:\epsilon^{\mu\nu\lambda}\\
&&\,\,\,\,\,\,\times\mathbf{tr}\left(G\partial_{\mu}G^{-1}G\partial_{\nu}G^{-1}G\partial_{\lambda}G^{-1}\right)\\
&=&\frac{1}{2\pi}\int_{0}^{2\pi}d\phi\int_{0}^{\frac{\pi}{2}}d\theta\;\frac{1}{|g|^3}\mathbf{g}\cdot\left(\partial_{\theta}\mathbf{g}\times\partial_{\phi}\mathbf{g}\right),
\end{eqnarray*}
with 
$$\mathbf{g}=\left(k\sin\phi\cos\theta\,,-k\cos\phi\cos\theta,k(\cos\phi+\sin\phi)+\Delta\sin\theta\right).$$
For simplicity, setting $\alpha=k=\Delta=1$, we obtain
\[
\nu[2,5]=\frac{1}{2\pi}\int_{0}^{2\pi}d\phi\int_{0}^{\frac{\pi}{2}}d\theta\,\,\, \frac{n(\theta,\phi)}{m(\theta,\phi)} =1,
\]
where
\[
n(\theta,\phi)=\cos\theta
\]
and
\begin{eqnarray*}
m(\theta,\phi)&=&[\sin^2\theta+\sin(2\theta)(\cos\phi+\sin\phi)\\
                      & &+\cos^2\theta(2+\sin(2\phi))]^{\frac{3}{2}}.
                      \end{eqnarray*}
It is then found that
a nontrivial $\mathbf{Z^{(1)}_{2}}$ topological charge $\nu(2,5)=1$, which equals to $N(4,5)$ obtained before.

\subsection{SII.B. Topological charge of the Fermi point on the boundary of quantum spin Hall systems}
For the boundary of a quantum spin Hall system, the effective boundary theory can be written as
\begin{equation}
\mathcal{H}(k_{x})=k_{x}\sigma_{1}+\alpha k_{x}\sigma_{2}+\beta k_{x}\sigma_{3},
\end{equation}
where the last two terms preserving TRS
will turn out to be irrelevant for the topological charge. When we choose a circle in the $(\omega,k_x)$-plane, the Green's function restricted on this circle is given by
\[
G^{-1}(\psi)=ia\sin\psi-\mathcal{H}(a\cos\psi)
\]
with $\psi\in[0,2\pi)$. Noting that TRS for the Green's function is $TG(\omega,k)T^\dagger=G^T(\omega,-k)$,  
this TRS is accordingly represented as $\sigma_2G(\psi)\sigma_2=G(\pi-\psi)$ on the circle. Then following Ref.\cite{FS-Classification}, we extend the Green's function restricted on the circle with two additional parameters, $i.e.$,
\begin{eqnarray*}
 &  & G^{-1}(\psi,\theta,\phi)\\
 & = & ia\sin\psi-\left[\left(\mathcal{H}(a\cos\psi)\cos\theta+\sin\theta\sigma_{2}\right)\cos\phi+\sin\phi\sigma_{3}\right]
\end{eqnarray*}
with $\theta\in[0,\pi]$ and $\phi\in[0,\pi]$. It is straightforward to verify that TRS is preserved for this extension:
\[
\sigma_{2}G\left(\psi,\theta,\phi\right)\sigma_{2}=G^{T}\left(\pi-\psi,-\theta,-\phi\right).
\]
 The $\mathbf{Z_2}^{(2)}$-type topological charge is calculated as 
\begin{eqnarray*}
\nu(1,5) & = & \frac{1}{24\pi^{2}}\int d\psi d\theta d\phi\epsilon^{\mu\nu\lambda}\\
 &  & \qquad\mathbf{tr}\left(G\partial_{\mu}G^{-1}G\partial_{\nu}G^{-1}G\partial_{\lambda}G^{-1}\right)\\
 & =& 1,
\end{eqnarray*}
which equals to $N(3,5)$ calculated before.

\section{SIII. Dirac Matrix Construction of FSs and TIs/TSCs and The Proof of General Index Theorem}

\subsection{SIII.A. Preliminaries of Dirac matrices}

The purpose of this section is to introduce the preliminaries about Dirac
matrices needed for our construction of topologically protected FSs and TIs/TSCs.
As for a pedagogical introduction to Dirac matrices, it is seen in Ref.\cite{Kim-Dirac-Matrices}.
Although most results may be basis-independent, we still introduce explicitly our convention of Dirac matrices below. 
\begin{eqnarray}
\Gamma_{(2n+1)}^{a} & = & \Gamma_{(2n-1)}^{a}\otimes\sigma_{1},\quad a=1,2,3,\cdots,2n-1,\nonumber \\
\Gamma_{(2n+1)}^{2n} & = & \mathbf{1}_{2^{n-1}}\otimes\sigma_{2},\nonumber \\
\Gamma_{(2n+1)}^{2n+1} & = & \mathbf{1}_{2^{n-1}}\otimes\sigma_{3},\label{eq:Dirac-Matrices}
\end{eqnarray}
where $\sigma_{i}$ are Pauli matrices and $\mathbf{1}_{2^{n-1}}$is
the $2^{n-1}\times2^{n-1}$unit matrix. We input the initial condition:
$\Gamma_{(3)}^{1}=\sigma_{1},\Gamma_{(3)}^{2}=\sigma_{2}$ and $\Gamma_{(3)}^{3}=\sigma_{3}$,
to obtain Dirac matrices in all dimensions. Notably $\Gamma$ matrices
with odd superscript are purely real, while ones with even superscript
are purely imaginary. All Dirac matrices are hermitian, and satisfy
the Clifford algebra: 
\begin{equation}
\left\{ \Gamma_{(2n+1)}^{a},\:\Gamma_{(2n+1)}^{b}\right\} =2\delta^{ab}\mathbf{1}_{2^{n}\times2^{n}}.\label{eq:Clifford-Algebra}
\end{equation}

We will construct Hamiltonians by Dirac matrices and discuss their
TRS and PHS. For this purpose,
we introduce the following operators.
\begin{eqnarray}
B_{(2n+1)}^{1} & := & \Gamma_{(2n+1)}^{3}\Gamma_{(2n+1)}^{5}\cdots\Gamma_{(2n+1)}^{2n+1}\nonumber \\
B_{(2n+1)}^{2} & := & \Gamma_{(2n+1)}^{2}\Gamma_{(2n+1)}^{4}\cdots\Gamma_{(2n+1)}^{2n}\nonumber \\
\tilde{B}_{(2n+1)}^{1} & := & B_{(2n+1)}^{1}\Gamma_{(2n+1)}^{2n+1}\label{eq:Symm-Matr}\\
\tilde{B}_{(2n+1)}^{2} & := & B_{(2n+1)}^{2}\Gamma_{(2n+1)}^{2n+1}\nonumber 
\end{eqnarray}
It is direct to verify the following commutation relations below:
\begin{eqnarray}
 &  & B_{(2n+1)}^{1}\Gamma_{(2n+1)}^{a}(B_{(2n+1)}^{1})^{-1}\nonumber \\
 & = & \begin{cases}
(-1)^{n+1}(\Gamma_{(2n+1)}^{a})^{T}, & a=2,\cdots,2n+1\nonumber\\
(-1)^{n}(\Gamma_{(2n+1)}^{1})^{T}\nonumber
\end{cases},\nonumber\\
 &  & B_{(2n+1)}^{2}\Gamma_{(2n+1)}^{a}(B_{(2n+1)}^{2})^{-1}\label{eq:Sym-Matrix}\\
 & = & (-1)^{n}(\Gamma_{(2n+1)}^{a})^{T},\quad a=1,2,\cdots,2n+1.\nonumber
\end{eqnarray}
The transposition matrices of $B^{1}$, $B^{2}$, $\tilde{B}^{1}$, and 
$\tilde{B}^{2}$ satisfy, respectively, 
\begin{eqnarray}
B_{(2n+1)}^{1} & = & (-1)^{n(n-1)/2}\left(B_{(2n+1)}^{1}\right)^{T}\nonumber \\
B_{(2n+1)}^{2} & = & (-1)^{n(n+1)/2}\left(B_{(2n+1)}^{2}\right)^{T}\nonumber \\
\tilde{B}_{(2n+1)}^{1} & =- & (-1)^{n(n+1)/2}\left(\tilde{B}_{(2n+1)}^{1}\right)^{T}\label{eq:Symm-Sign}\\
\tilde{B}_{(2n+1)}^{2} & = & (-1)^{n(n-1)/2}\left(\tilde{B}_{(2n+1)}^{2}\right)^{T}\nonumber
\end{eqnarray}

\subsection{SIII.B. Construction of all topological types of Fermi surfaces}

\subsubsection{Fermi surfaces in non-chiral classes AI, AII, D and C}

$\mathbf{Z}\,\,and\,\,\mathbf{Z}_{2}^{(1,2)}$--The simplest way to construct a Hamiltonian with Dirac matrices is
to make Weyl-type Hamiltonian, namely,
\begin{equation}
\mathcal{H}_{(2n+1)}^{W}(\mathbf{k})=\sum_{a=1}^{2n+1}k_{a}\Gamma_{(2n+1)}^{a}.\label{eq:Weyl-type}
\end{equation}
This Hamiltonian has either TRS or PHS depending on its spatial dimension $2n+1$. To be explicit, the Hamiltonian satisfies
\begin{eqnarray}
 &  & B_{(2n+1)}^{2}\mathcal{H}_{(2n+1)}^{W}(\mathbf{k})(B_{(2n+1)}^{2})^{-1}\nonumber\\
 & = & (-1)^{n+1}\left(\mathcal{H}_{(2n+1)}^{W}\right)^{T}(-\mathbf{k}), \label{eq:Weyl-Symm-Oper}
\end{eqnarray}
which can be seen from Eq.(\ref{eq:Symm-Sign}). $B^{2}$ represents
a TRS with $n=2m+1$ or a PHS with $n=2m$, where $m$ is an integer or zero,
and the corresponding symmetry sign is given by Eq.(\ref{eq:Symm-Sign}).
This Hamiltonian has a Fermi point at the origin of $\mathbf{k}$-space,
whose codimension is $p=2n+1$.
The Fermi point is in classes AI, AII, D, and C, respectively, for $p=8m+7$,  $p=8m+3$,  $p=8m+1$,
and  $p=8m+5$, which precisely coincides 
with that of $\mathbf{Z}$-type topological charge for the same class
in the classification table of Fermi surfaces (Tab.[I] in the main text).

Our next task is to show that this Fermi point has a unit $\mathbf{Z}$-type
topological charge. It can be enclosed by a cylinder $A:=(-\infty,\infty)\times S^{2n}$
in $(\omega,\mathbf{k})$-space with $S^{2n}$ centered at the origin
of $\mathbf{k}$-space and $\omega$ being in $(-\infty,\infty)$.
We parametrize the Green's function on this cylinder $A$ as 
\[
G^{-1}|_{A}(\omega,\mathbf{k})=i\omega-k_{a}(s)\Gamma_{(2n+1)}^{a},
\]
where $k_a(s)$ are the same as those defined in Eq.(\ref{Spherical-Coordinates}) with $p=2n+1$. For the Green's
function, the cylinder $A$ is essentially a topological $(2n+1)$-dimensional
sphere $S^{2n+1}$ since $G^{-1}$ tends to a constant matrix when
$\omega$ approaches to $\pm\infty$. To compute its topological charge,
we make the following integral on the cylinder with Eq.(\ref{eq:Clifford-Algebra}).
\begin{eqnarray}
\nu & = & C_{2n+1}\int_{A}\left(G\mathbf{d}G\right)^{2n+1}\nonumber \\
 & = & (2n+1)C_{2n+1}\int_{-\infty}^{\infty}d\omega\int_{S^{2n}}d^{2n}s\;\epsilon^{\mu_{1}\mu_{2}\cdots\mu_{2n}}\nonumber \\
 &  & \qquad\qquad\mathbf{tr}\left(G\partial_{\omega}G^{-1}G\partial_{\mu_{1}}G^{-1}\cdots G\partial_{\mu_{2n}}G^{-1}\right)\nonumber \\
 & = & \frac{1}{2^{n-1}\Omega_{2n+1}}\int_{S^{2n}}d^{2n}s\;\frac{1}{|\mathbf{n}|^{2n+1}}\epsilon^{\mu_{1}\mu_{2}\cdots\mu_{2n+1}}\epsilon^{\nu_{1}\nu_{2}\cdots\nu_{2n}}\nonumber \\
 &  & \qquad\qquad n_{\mu_{1}}\partial_{\nu_{1}}n_{\mu_{2}}\partial_{\nu_{2}}n_{\mu_{3}}\cdots\partial_{\nu_{2n}}n_{\mu_{2n+1}}\nonumber \\
 & = & 1\label{eq:Weyl-Charge}
\end{eqnarray}
with $\Omega_{2n+1}$ being the total solid angle of $(2n+1)$-dimensional
space. Now it is clear that all Fermi points of unit topological charge
with either a TRS or PHS can be constructed by the Dirac matrices
in the form of Eq.(\ref{eq:Weyl-type}). Note that we can construct an $N$-charged Fermi point for each case simply by mutipling the unit Hamiltonian $N$ times for $\mathbf{Z}$ types, or the double-unit models (to be constructed latter) $N/2$ times for $2\mathbf{Z}$ types. This also holds for $\mathbf{Z}$-type and $2\mathbf{Z}$-type TIs/TSCs with multiple topological numbers, which will not be repeated any more.  

We now turn to construct FSs protected by $\mathbf{Z}_{2}^{(1,2)}$
type topological charge. According to our general classification table,
each $\mathbf{Z}$ type topological charge is followed by $\mathbf{Z}_{2}^{(1)}$
and $\mathbf{Z}_{2}^{(2)}$ with decreasing codimensions, which motivates
us to construct corresponding models by lowering spatial dimensions,
i.e. the number of $k$s from the Weyl-type model (\ref{eq:Weyl-type}).
For $\mathbf{Z}_{2}^{(1)}$ cases, we consider the following model
\begin{equation}
\mathcal{H}_{W}^{(1)}=\sum_{a=1}^{2n}k_{a}\Gamma^{a+1}+\lambda\sum_{a=1}^{2n}k_{a}\Gamma^{1},\label{eq:First-Weyl}
\end{equation}
where the last term with a coefficient $\lambda$ is added to merely
indicate that the considered model, while keeping TRS or PHS with the topological character unchanged, may not need
to have an additional CS. We may set $\lambda=0$ without changing the topological
property of the Fermi point at $k=0$. For brevity,  we have
dropped the awkward subscript $(2n+1)$ in the above expression and will do it hereafter
 if it can be recovered from the context. To calculate
the $\mathbf{Z}_{2}^{(1)}$ type topological charge of the Fermi point,
we first choose an $S^{2n-1}$ in $\mathbf{k}$ space, which is parametrized
by the spherical coordinates $s_{i}$ with $i=1,2,\cdots,2n-1$, and
restrict the Hamiltonian on it, i.e. $\mathcal{H}_{W}^{(2)}|_{S^{2n-1}}=\mathcal{H}_{W}^{(2)}(s)$.
According to the formulation of $\mathbf{Z}_{2}^{(1)}$ type topological
charge, we make a one-parameter extension of $\mathcal{H}_{W}^{(2)}(s)$,
which may be chosen as 
\[
\mathcal{H}(s,\alpha)=\mathcal{H}(s)\cos\alpha+\sin\alpha\Gamma^{1},
\]
with $\alpha\in[-\pi/2,\pi/2]$. It is straightforward to check that
this extension satisfies all the requirements\cite{FS-Classification}. Using the extended
Green's function $G^{-1}(\omega,s,\alpha)=i\omega-\mathcal{H}(s,\alpha)$,
we compute the topological charge as 
\[
\nu=C_{2n+1}\int_{\tilde{A}}\mathbf{tr}\left(G\mathbf{d}G^{-1}(\omega,s,\alpha)\right)^{2n+1},
\]
where $\tilde{A}=A\times S^{1}$ with $A$ being the original domain
$(-\infty,\infty)\times S^{2n-1}$ in $(\omega,\mathbf{k})$ space.
After replacing the variable $\alpha$ with $\alpha'=\alpha+\pi/2$,
we are able to treat $\alpha$ as the $2n$-th spherical coordinate
of $S^{2n}$ made by $S^{2n-1}$ and the extended dimension. Thus
referring to our calculation of $\mathbf{Z}$ type topological charge,
we find that $\nu=1\mod2$, which means that the Fermi point at $k=0$
for model (\ref{eq:First-Weyl}) is nontrivial.

With the experience of constructing nontrivial $\mathbf{Z}_{2}^{(1)}$
type topological charge, we now give the following model as a candidate
for $\mathbf{Z}_{2}^{(2)}$:
\begin{eqnarray}
\mathcal{H}_{W}^{(2)} & = & \sum_{a=1}^{2n-1}k_{a}\Gamma^{a+2}\nonumber\\
 &  & +\lambda_{1}\sum_{a=1}^{2n-1}k_{a}\Gamma^{1}+\lambda_{2}\sum_{a=1}^{2n-1}k_{a}^{3}\Gamma^{2},\label{eq:Second-Weyl}
\end{eqnarray}
where analogous to the previous case, terms with $\lambda_{1,2}$
are added to mean that the model does not require additional
symmetries, and again they may be set as zero without changing
the topological property of the Fermi point. To see the topological
property of the Fermi point, we choose an $S^{2n-2}$, which is parametrized
with a spherical coordinates $s_{i}$ with $i=1,2,\cdots,2n-2$, in
$\mathbf{k}$ space to enclose it. The Hamiltonian restricted on the
$S^{2n-2}$ is denoted by $\mathcal{H}_{W}^{(2)}(s)$, and then according
to the formulation of $\mathbf{Z}_{2}^{(2)}$ type topological number,
we make two-parameter extension of $\mathcal{H}_{W}^{(2)}(s)$, which
may be written as 
\[
\mathcal{H}_{W}^{(2)}(s,\alpha,\beta)=\left(\mathcal{H}_{W}^{(2)}(s)\cos\alpha+\sin\alpha\Gamma^{1}\right)\cos\beta+\sin\beta\Gamma^{2},
\]
where $\alpha,\beta\in[-\pi/2,\pi/2]$. The above extension  satisfies
all the requirements for computing the topological charge
of the Fermi point\cite{FS-Classification}. Using the extended Green's function $G^{-1}(\omega,s,\alpha,\beta)=i\omega-\mathcal{H}_{W}^{(2)}(s,\alpha,\beta)$,
we have the topological charge as 
\[
\nu=C_{2n+1}\int_{\tilde{\tilde{A}}}\mathbf{tr}\left(G\mathbf{d}G^{-1}(\omega,s,\alpha,\beta)\right),
\]
where $\tilde{\tilde{A}}=A\times T^{2}$ with $A=[-\infty,\infty]\times S^{2n-2}$
and $T^{2}$ being the domain of $\alpha$ and $\beta$. Through the
variable transformation: $\alpha'=\alpha+\pi/2$ and $\beta'=\beta+\pi/2$,
we can combine $T^{2}$ and $S^{2n-2}$ as $S^{2n}$. Thus the Fermi
point has nontrivial topological charge $\nu=1\mod2$.

$2\mathbf{Z}\,\,type$--To construct $2\mathbf{Z}$ type Fermi surface in our classification
table, we introduce the following operator:
\[
\mathcal{B}^{1}=\Gamma^{2}\Gamma^{5}\Gamma^{7}\cdots\Gamma^{2n+1},
\]
which is a modified version of $B^{1}$ with $\Gamma^{3}$ being
replaced by $\Gamma^{2}$. $\mathcal{B}^{1}$ satisfies
\begin{equation}
\mathcal{B}^{1}=-(-1)^{n(n-1)/2}\left(\mathcal{B}^{1}\right)^{T}\label{eq:Sign-Of-MB1}
\end{equation}
and 
\begin{equation}
\begin{cases}
\mathcal{B}^{1}\Gamma^{a}\left(\mathcal{B}^{1}\right)^{-1}=(-1)^{n+1}\left(\Gamma^{a}\right)^{T} & 4\leq a\leq2n+1\\
\mathcal{B}^{1}\tilde{\Gamma}\left(\mathcal{B}^{1}\right)^{-1}=\left(-1\right)^{n+1}\tilde{\Gamma}^{T}
\end{cases}\label{eq:M-B1-Relation}
\end{equation}
with $\tilde{\Gamma}=i\Gamma^{1}\Gamma^{2}\Gamma^{3}$. We construct
the model as
\begin{equation}
\mathcal{H}_{W}^{2}(\mathbf{k})=\sum_{a=2}^{2n-1}k_{a}\Gamma_{(2n+1)}^{a+2}+\tilde{\Gamma}_{(2n+1)}k_{1}.\label{eq:FS-2Z-nonchiral}
\end{equation}
As $\mathcal{H}_{W}^{2}$ can be diagonalized into two equivalent
$2^{n-1}\times2^{n-1}$ blocks since $\left[\mathcal{H}_{W}^{2},i\Gamma^{1}\Gamma^{2}\right]=0$,
up to a unitary transformation it can rewritten as
\[
\mathcal{H}_{W}^{2}(\mathbf{k})\sim\begin{pmatrix}\sum_{a=1}^{2n-1}k_{a}\Gamma_{(2n-1)}^{a}\\
 & \sum_{a=1}^{2n-1}k_{a}\Gamma_{(2n-1)}^{a}
\end{pmatrix}.
\]
Since each block has unit $\mathbf{Z}$ type topological charge, the
Fermi point of $\mathcal{H}(\mathbf{k})$ has topological charge $2$.
We then analyze its TRS and PHS by using Eq.(\ref{eq:M-B1-Relation}),
which leads to
\[
\mathcal{B}^{1}\mathcal{H}_{W}^{2}(\mathbf{k})\left(\mathcal{B}^{1}\right)^{-1}=(-1)^{n}\mathcal{H}_{W}^{2T}(-\mathbf{k}).
\]
Thus based on the relation Eq.(\ref{eq:Sign-Of-MB1}) and the above relation, we summarize that the Fermi point  is in class AI, AII, D, and C, respectively, for the codimension
$p=8m+3$, $p=8m+7$, $p=8m+5$,
and $p=8m+1$.

\subsubsection{Fermi surfaces in chiral classes CI, CII, DIII and BDI}

$\mathbf{Z}\,\,\,type\,\,\,and\,\,\,\mathbf{Z}_{2}^{(1,2)}\,\,\,type$-To handle the remaining four symmetry classes with chiral symmetry(CS),
we construct the Dirac-type Hamiltonian as
\begin{equation}
\mathcal{H}_{D}(\mathbf{k})=\sum_{a=1}^{2n}k_{a}\Gamma^{a}.\label{eq:Dirac-type}
\end{equation}
In other words, we sum over all $\Gamma$s except the last one, which
leads to CS, $i.e.$
\[
\left\{ \mathcal{H}_{D},\Gamma^{2n+1}\right\} =0.
\]
As is known that this Hamiltonian has a TRS(PHS) represented
by $B^{2}$ just like that of $\mathcal{H}^{W}$, using $\Gamma_{(2n+1)}^{2n+1}$
we are able to newly construct a PHS(TRS), which is represented by
$\tilde{B}^{2}=B^{2}\Gamma^{2n+1}$ as in Eq.(\ref{eq:Symm-Matr}).
It is verified that
\begin{eqnarray*}
\tilde{B}^{2}\mathcal{H}_{D}(\mathbf{k})\left(\tilde{B}^{2}\right)^{-1} & = & -B^{2}\mathcal{H}_{D}(\mathbf{k})\left(B^{2}\right)^{-1}\\
 & = & (-1)^{n}\left(\mathcal{H}_{D}(-\mathbf{k})\right)^{T}
\end{eqnarray*}
Compared with Eq.(\ref{eq:Weyl-Symm-Oper}), if $B^{2}$ refers to
a TRS, then $\tilde{B}^{2}$ refers to a PHS, and vice versa. The
sign of $\tilde{B}^{2}$ as a TRS/PHS can be seen from Eq.(\ref{eq:Symm-Sign}).
As the codimension of the Fermi point for the Hamiltonian of Eq.(\ref{eq:Dirac-type})
is $2n$, the Fermi point is in class CI, DIII, BDI, and CII respectively when $p=8m+6$, $p=8m+2$, $p=8m$, and $p=8m+4$, which is in agreement with the appearance of $\mathbf{Z}$-type
topological charge for the four classes with CS in the classification
table. Parallel to the case without CS, we will show that the Fermi
point of Eq.(\ref{eq:Dirac-type}) has unit charge. Before doing it,
it is noted that $\mathcal{H}_{D}$ can be expressed as 
\begin{equation}
\mathcal{H}_{D}=\begin{pmatrix}0 & u\\
u^{\dagger} & 0
\end{pmatrix}, \label{eq:Off-Diagonal}
\end{equation}
where $u=-ik_{2n}\mathbf{1}_{2^{n-1}\times2^{n-1}}+\sum_{a=1}^{2n-1}k_{a}\Gamma_{(2n-1)}^{a}$
with our convention of Dirac matrices in Eq.(\ref{eq:Dirac-Matrices}).
To calculate the topological charge, we choose an $S^{2n-1}$ in the
$\mathbf{k}$-space to enclose the Fermi point, which is parametrized
by $s_{i}$ with $i$ being from $1$ to $2n-1$. Thus we have 
\begin{eqnarray*}
\nu & = & \frac{C_{2n-1}}{2}\int_{S^{2n-1}}\mathbf{tr}\left[\Gamma^{2n+1}\left(\left(\mathcal{H}_{D}\right)^{-1}\mathbf{d}\mathcal{H}_{D}\right)^{2n-1}\right].
\end{eqnarray*}
Without loss of generality, we choose the $S^{2n-1}$ to be unit,
which implies $u^{\dagger}u=1$ and leads to
\[
\nu=C_{2n-1}\int_{S^{2n-1}}\mathbf{tr}\left(u\mathbf{d}u^{\dagger}\right)^{2n-1}.
\]
As $u^{\dagger}$ can be regarded as an inversion of Green's function
in $(2n-1)$ spatial dimensions with $k_{2n}$ as $\omega$, referring
to the case of Eq.(\ref{eq:Weyl-Charge}), it is seen that $\nu=1$. 

Analogous to the cases without CS, we can construct $\mathbf{Z}_{2}^{(1,2)}$
type topological Fermi surfaces by lowering spatial dimension from
the model (\ref{eq:Dirac-type}). The process is entirely analogous
to that of non-chiral cases, so we simply write down the corresponding
models below:
\begin{equation}
\begin{cases}
\mathcal{H}_{D}^{(1)}=\sum_{a=1}^{2n-1}k_{a}\Gamma^{a+1} & for\quad\mathbf{Z}_{2}^{(1)}\\
\mathcal{H}_{D}^{(2)}=\sum_{a=1}^{2n-2}k_{a}\Gamma^{a+2} & for\quad\mathbf{Z}_{2}^{(2)}
\end{cases}.\label{eq:FS-Dirac-Z2}
\end{equation}
Note that although the above two models  have additional symmetries,
e.g., $\Gamma^{1}$ anti-commutes with $\mathcal{H}_{D}^{(1)}$, these
symmetries are not required, which means that we can add some terms to break
them without changing the corresponding topological properties, such
as the $\lambda$-terms in model (\ref{eq:First-Weyl})
and model (\ref{eq:Second-Weyl}).

$2\mathbf{Z}\,\,\,type$-According to the corresponding model of classes AI, AII, C and D,
we can define
\[
\tilde{\mathcal{B}}^{1}:=\mathcal{B}^{1}\Gamma^{2n+1},
\]
which satisfies the relation
\begin{equation}
\tilde{\mathcal{B}}^{1}=(-1)^{n(n+1)/2}\left(\tilde{\mathcal{B}}^{1}\right)^{T}.\label{eq:Sign-of-VMB1}
\end{equation}
Accordingly, we introduce the following model:
\begin{equation}
\mathcal{H}_{D}^{2}(\mathbf{k})=\sum_{a=2}^{2n-2}k_{a}\Gamma^{a+2}+k_{1}\tilde{\Gamma},\label{eq:FS-Dirac-2Z}
\end{equation}
whose Fermi point has a topological charge $2$ from the same reasoning
for model (\ref{eq:FS-2Z-nonchiral}). It has a CS (represented by $\Gamma^{2n+1}$);
and moreover, when $\mathcal{B}^{1}$ represents a TRS(PHS), $\tilde{\mathcal{B}}^{1}$
denotes a PHS(TRS). Considering the relation (\ref{eq:Sign-of-VMB1})
and the results of nonchiral cases, it is found that the Fermi point
is in classes DIII, CI, CII, and BDI, respectively, when $p=8m+6$, $p=8m+2$, $p=8m$, and 
$p=8m+4$, which is in consistence with our classification table of
Fermi surfaces.

To conclude this subsection, all types of Fermi surfaces
in the classification table have been constructed by Dirac matrices.

\subsection{SIII.C. Construction of all topological types of TIs/TSCs}

\subsubsection{TIs/TSCs in non-chiral classes AI, AII, D and C}

$\mathbf{Z}\,\,\,type\,\,\,and\,\,\,\mathbf{Z}_{2}^{(1,2)}\,type$--
We first wish to indicate that a part of results to be presented for the relationship of
Dirac matrices with TIs/TSCs were addressed in Ref.\cite{Ryu-TI-Classification}.
As the bulk energy spectrum of a TI/TSC is fully gaped, we need one of $\Gamma$s
with a coefficient as a mass term, which is chosen to be $\Gamma^{1}$.
Thus a typical Hamiltonian reads 
\begin{equation}
\mathcal{H}(\mathbf{k})=\sum_{a=2}^{2n+1}k_{a-1}\Gamma^{a}+\left(m-\epsilon k^{2}\right)\Gamma^{1}.\label{eq:Modify-Dirac}
\end{equation}
Note that the term of $\epsilon k^{2}\Gamma^{1}$, in which a constant $\epsilon$
can be regarded to be infinitesimally small, is included
as a prescription for the singularity of this kind of continuum model at the infinity
in the $\mathbf{k}$ space\cite{Note-Lattice}. It will turn out  that the topological number of this Hamiltonian is either $\pm1$ or $0$. 

Let us look into the
TRS/PHS of Hamiltonian in Eq.(\ref{eq:Modify-Dirac}), which is
now represented by $B^{1}$. Using Eq.(\ref{eq:Sym-Matrix}), we 
obtain the relation
\begin{equation}
B^{1}\mathcal{H}(\mathbf{k})(B^{1})^{-1}=(-1)^{n}\mathcal{H}^{T}(-\mathbf{k}).\label{eq:TRS/PHS-TI}
\end{equation}
Thus when $n=2m$, $i.e.$ the spatial dimension $d=4m$, $B^{1}$
represents a TRS, while it corresponds to a PHS when $n=2m+1$ or $d=4m+2$.
Eq.(\ref{eq:Symm-Sign}) can tell us the sign of the TRS/PHS, so that
we can encapsulate that Hamiltonian of Eq.(\ref{eq:Modify-Dirac})
is in classes AI, AII, D, and C, respectively, for $d=8m$, $d=8m+4$, $d=8m+2$, $d=8m+6$, which is in agreement with the distribution
of $\mathbf{Z}$-type topological numbers in the classification of
TIs/TSCs. The remaining task for these classes is to verify that 
Hamiltonian of Eq.(\ref{eq:Modify-Dirac}) has indeed unit
topological number. Substituting Eq.(\ref{eq:Modify-Dirac})
into the corresponding formula for the integer topological number, i.e., 
\[
N=C_{d+1}\int_{\mathcal{M}}\mathbf{tr}\left(G\mathbf{d}G^{-1}(\omega,\mathbf{k})\right)^{d+1},
\]
where $\mathcal{M}$ is the whole $(\omega,\mathbf{k})$-space, we
can obtain
\begin{equation}
N=\frac{1}{2}\left(\mathbf{sgn}(m)+\mathbf{sgn}(\epsilon)\right),\label{eq:TN-Of-MD}
\end{equation}
which indicates unit topological number when
both $m$ and $\epsilon$ have the same sign. Thus it is clear that the model
of Eq.(\ref{eq:Modify-Dirac}) can be regarded as a representative
for each $\mathbf{Z}$-type of the classes AI, AII, D and C.

Since the procedures to construct the $\mathbf{Z}_{2}^{(1,2)}$ type TIs/TSCs are
similar to those for constructing $\mathbf{Z}_{2}^{(1,2)}$ type Fermi
surfaces, to avoid redundant derivations,
we here write
down the results directly with some necessary remarks.
The corresponding models for cases of $\mathbf{Z}_{2}^{(1)}$ and
$\mathbf{Z}_{2}^{(1)}$ are, respectively,
\[
\begin{cases}
\mathcal{H}^{(1)}=\sum_{a=1}^{2n-1}k_{a}\Gamma^{a+2}+\left(m-\epsilon k^{2}\right)\Gamma^{1} & for\;\mathbf{Z}_{2}^{(1)}\\
\mathcal{H}^{(2)}=\sum_{a=1}^{2n-2}k_{a}\Gamma^{a+3}+\left(m-\epsilon k^{2}\right)\Gamma^{1} & for\;\mathbf{Z}_{2}^{(2)}
\end{cases}.
\]
To see that nontrivial topological property of $\mathcal{H}^{(1,2)}$,
we need to make continuous extension with new parameters, which can
be chosen as the omitted momenta, compared with model (\ref{eq:Modify-Dirac}).
Concretely, for $\mathcal{H}^{(1)}$ the extension can be made by
adding the term $\alpha\Gamma^{2}-\epsilon\alpha^{2}\Gamma^{1}$ with
$\alpha\in(-\infty,\infty)$. When $\alpha=0$, we obtain the original
$\mathcal{H}^{(1)}$, and by changing $\alpha$ from $0$ to $\pm\infty$,
$\mathcal{H}^{(1)}$ is deformed smoothly without closing its gap
to a trivial model $\sim\alpha\Gamma^{2}-\epsilon\alpha^{2}\Gamma^{1}$.
Since the extended model is equivalent to the model (\ref{eq:Modify-Dirac})
whose $\mathbf{Z}$ type topological number is given by Eq.(\ref{eq:TN-Of-MD}),
it is concluded that the $\mathbf{Z}_{2}^{(1)}$ type topological number of
$\mathcal{H}^{(1)}$ is also given by Eq.(\ref{eq:TN-Of-MD}). The
same reasoning is applicable to $\mathcal{H}^{(2)}$ when we make
two-parameter extension, thus $\mathcal{H}^{(2)}$ has a nontrivial
$\mathbf{Z}_{2}^{(2)}$ type topological number when $m$ and $\epsilon$
have the same sign. We note again that only the symmetries in the
corresponding symmetry class are required, while other terms preserving
topological properties may be added to break unwanted additional symmetries.

$2\mathbf{Z}\,\,\,type$--Similar to the case of $2\mathbf{Z}$ type Fermi surface, we
introduce a model as
\[
\mathcal{H}^{2}(\mathbf{k})=\sum_{a=1}^{2n-2}k_{a}\Gamma^{a+3}+\left(m-\epsilon k^{2}\right)\tilde{\Gamma}
\]
with topological number $\mathbf{sgn}(m)+\mathbf{sgn}(\epsilon)$,
which can be seen from our previous discussion of $2\mathbf{Z}$ type
Fermi surfaces. Noting that
\[
B^{2}\tilde{\Gamma}\left(B^{2}\right)^{-1}=\left(-1\right)^{n+1}\tilde{\Gamma}
\]
and relation (\ref{eq:Sym-Matrix}), we have
\begin{equation}
B^{2}\mathcal{H}^{2}(\mathbf{k})\left(B^{2}\right)^{-1}=\left(-1\right)^{n+1}\mathcal{H}^{2T}(-\mathbf{k}).\label{eq:Sym-2Z-TI-Nonchiral}
\end{equation}
Combining Eq.(\ref{eq:Symm-Sign}) with Eq.(\ref{eq:Sym-2Z-TI-Nonchiral}),
it can be found that the classes of model belong, respectively, to  AI, AII, D, and C for $d=8m+4$, $d=8m$, $d=8m+6$, and $d=8m+2$,
consistent with our classification of TIs/TSCs.

\subsubsection{TIs/TSCs in chiral classes CI, CII, BDI and DIII}

$\mathbf{Z}\,\,\,type\,\,\,and\,\,\,\mathbf{Z}_{2}^{(1,2)}\,type$--For the other classes with CS, the term $\Gamma^{2n+1}$ is excluded for the presence of
CS, and the corresponding model may be written as
\begin{equation}
\mathcal{H}_{CS}(\mathbf{k})=\sum_{a=2}^{2n}k_{a-1}\Gamma^{a}+\left(m-\epsilon k^{2}\right)\Gamma^{1}.\label{eq:CS-M-Dirac}
\end{equation}
Here the additional TRS/PHS is represented by $\tilde{B}^{1}=B^{1}\Gamma^{2n+1}$
referring to Eq.(\ref{eq:Symm-Matr}), and we can check that
\begin{eqnarray*}
\tilde{B}^{1}\mathcal{H}_{CS}(\mathbf{k})(\tilde{B}^{1})^{-1} & = & -B^{1}\mathcal{H}_{CS}(\mathbf{k})(B^{1})^{-1}\\
 & = & (-1)^{n+1}\mathcal{H}_{CS}^{T}(-\mathbf{k}).
\end{eqnarray*}
Compared with Eq.(\ref{eq:TRS/PHS-TI}), it is seen that if $B^{1}$
denotes a TRS(PHS), then $\tilde{B}^{1}$ corresponds to a
PHS(TRS). The sign of $\tilde{B}^{1}$ is given by Eq.(\ref{eq:Symm-Sign}),
and thus it can be seen that the classes of model (\ref{eq:CS-M-Dirac}) belong, respectively, to
 CI, DIII, BDI, and CII for $d=8m+7$, $d=8m+3$, $d=8m+1$, and $d=8m+5$. To check the $\mathbf{Z}$-type
topological number, we use the corresponding formula
\[
N=\frac{C_{2n-1}}{2}\int_{M}\mathbf{tr}\left[\Gamma^{2n+1}\left(\mathcal{H}_{CS}^{-1}\mathbf{d}\mathcal{H}_{CS}\right)^{2n-1}\right],
\]
where $M$ denotes the whole $\mathbf{k}$ space. If we recover the
$\Gamma$ matrices with their subscript to be $\Gamma_{(2n+1)}^{a}$, we
recall Eq.(\ref{eq:Off-Diagonal}) of $\mathcal{H}_{CS}$ with
\begin{eqnarray*}
u & = & -ik_{2n-1}\mathbf{1}_{2^{n-2}}\\
 &  & +\sum_{a=2}^{2n-1}k_{a-1}\Gamma_{(2n-1)}^{a}+(m-\epsilon k^{2})\Gamma_{(2n-1)}^{1}.
\end{eqnarray*}
Similar to the calculation of topological charge for Fermi point,
we have
\[
N=C_{2n-1}\int_{M}\mathbf{tr}\left(u\mathbf{d}u^{\dagger}\right)^{2n-1}.
\]
Thus the topological number is given by
\[
N=\frac{1}{2}(\mathbf{sgn}(m)+\mathbf{sgn}(\epsilon)),
\]
which implies $N=1$ when both $m$ and $\epsilon$ are positive.
Note that the correct definition of $k^{2}$ should be $k^{2}=\sum_{a=1}^{2n-2}k_{a}^{2}$
in order to obtain the above expression , since adding $k_{2n-1}^{2}$
in the summation will always make the topological number vanish. 

$\mathbf{Z}_{2}^{(1,2)}$ type TIs/TSCs in these classes can be obtained
directly by the method used in the previous section, thus we here 
present the result:
\[
\begin{cases}
\mathcal{H}_{CS}^{(1)}=\sum_{a=1}^{2n-2}k_{a}\Gamma^{a+2}+\left(m-\epsilon k^{2}\right)\Gamma^{1} & for\;\mathbf{Z}_{2}^{(1)}\\
\mathcal{H}_{CS}^{(2)}=\sum_{a=1}^{2n-3}k_{a}\Gamma^{a+3}+\left(m-\epsilon k^{2}\right)\Gamma^{1} & for\;\mathbf{Z}_{2}^{(2)}
\end{cases}
\]
The two models have nontrivial $\mathbf{Z}_{2}^{(1,2)}$ type
topological numbers when $m$ and $\epsilon$ have the same sign.

\textit{$2\mathbf{Z}$ type}--From our experience, a desired mode in this case can be written as
\[
\mathcal{H}_{CS}^{2}(\mathbf{k})=\sum_{a=1}^{2n-3}k_{a}\Gamma^{a+3}+\left(m-\epsilon k^{2}\right)\tilde{\Gamma},
\]
which has topological number $\mathbf{sgn}(m)+\mathbf{sgn}(\epsilon)$.
If $B^{2}$ corresponds to a TRS(PHS), $\tilde{B}^{2}$ represents
a PHS(TRS). Then the classes of the above model belong, respectively, to DIII, CI, BDI, and CII when $d=8m+7$, $d=8m+3$, $d=8m+5$, and 
$d=8m+1$, in agreement with our classification table of TIs/TSCs.

To conclude this subsection, all types of TIs/TSCs in our classification
table have been constructed by using Dirac matrices.

\subsection{SIII.D. Boundary Modes of Dirac Models and General Index Theorem}

Up to now, all types of TIs/TSCs have been constructed by Dirac matrices,
and all of our constructed models may be written in the unified form:
\begin{equation}
\mathcal{H}_{G}(\mathbf{k})=\sum_{a=1}^{d}k_{a}\Gamma^{a+b-1}+\left(m-\epsilon k^{2}\right)\Gamma^{\alpha},\label{eq:Unified-TI}
\end{equation}
where $d$ is the bulk dimension, $b$ is the starting superscript
of $\Gamma_{(2n+1)}$ matrices, and $\Gamma^{\alpha}=\Gamma^{1}$
except in the cases of $2\mathbf{Z}$-type where $\Gamma^{\alpha}=\tilde{\Gamma}$.
The advantage of this kind of construction in the form of (\ref{eq:Unified-TI})
lies in that its boundary low-energy effective theory can be formulated
systematically through the perturbation theory of quantum mechanics. The method used here is a generalization of that in Ref.\cite{SQS-Book}.
We consider that a boundary at $x=0$ is on the left of a $d$-dimensional
model of \ref{eq:Modify-Dirac} and translation invariance is still
preserved along the other $d-1$ directions. To implement the perturbation
method, we will first identify the gapless subspace of the model residing
on the boundary, i.e., concentrated near $x=0$, and then compute
the transition elements in this subspace by regarding the remaining
translation invariant terms as perturbations. 

A physical boundary state $\varphi(\mathbf{x})$ has zero energy
and should vanish at $x=0$ and $x\rightarrow+\infty$. If such
a state exists, it satisfies the following equation
\[
\left[-i\Gamma^{b}\partial_{x}+\left(m+\epsilon\partial_{x}^{2}\right)\Gamma^{1}\right]\varphi(x)=0,
\]
where momenta along the other directions are set to be zero since only the ground state is relevant at present.
 Also note  that we have set $\Gamma^{\alpha}=\Gamma^{1}$
for explicitness, since there is no difference essential for the cases of $\tilde{\Gamma}$. The above equation can be rewritten as 
\[
\left[\partial_{x}+\left(m+\epsilon\partial_{x}^{2}\right)i\Gamma^{b}\Gamma^{1}\right]\varphi(x)=0.
\]
Assuming that $\varphi(x)=\chi_{\eta}f(x)$, where $\eta=\pm$ and
$\chi_{\pm}$ is the eigenvector of $i\Gamma^{b}\Gamma^{1}$ with
$\pm$ being the corresponding eigenvalue. Then 
\[
\partial_{x}f(x)+\eta\left(m+\epsilon\partial_{x}^{2}\right)f(x)=0,
\]
with the boundary conditions:
\[
f(0)=0\qquad and\qquad f(x)|_{x\rightarrow\infty}=0.
\]
Seeking solutions with the form $f\sim e^{-\lambda x}$, we have 
\[
\lambda^{2}-\frac{\eta}{\epsilon}\lambda+\frac{m}{\epsilon}=0.
\]
To satisfy the boundary condition we need the two roots $\lambda_{1,2}$
are both positive, which requires $\eta=\mathbf{sgn}(\epsilon)$ and
$m\epsilon>0$. Thus the condition of the existence of the boundary
states is 
\begin{equation}
\mathbf{sgn}(m)=\mathbf{sgn}(\epsilon),\label{eq:Existence-Condition}
\end{equation}
which is in consistence with the bulk topological number, $e.g.$,
expressed in Eq.(\ref{eq:TN-Of-MD}). It turns out that there exist
$2^{n-1}$ degenerate solutions with degeneracy originated from the
internal space:
\[
\varphi_{i}(x)=\chi_{\mathbf{sgn}(\epsilon)}^{i}\left(e^{-\lambda_{1}x}-e^{-\lambda_{2}x}\right),
\]
where $i$ labels the $2^{n-1}$ degenerate eigenvectors of $i\Gamma^{b}\Gamma^{1}$
with eigenvalue $\mathbf{sgn}(\epsilon)$. 

To obtain the low-energy effective Hamiltonian on the boundary, we
consider the remaining terms along the other directions of Eq.(\ref{eq:Modify-Dirac}):
\[
\Delta\mathcal{H}=\sum_{a=2}^{d}k_{a}\Gamma^{a+b-1}-\epsilon\sum_{a=2}^{d}k_{a}^{2}\Gamma^{1}
\]
as perturbations. Since we are only interested in the low-energy behavior
of the boundary, it is sufficient to implement the perturbation theory
of quantum mechanics in this subspace with zero energy to obtain the
low-energy effective Hamiltonian $\mathcal{H}_{eff}$, that is 
\[
\mathcal{H}_{eff}^{ij}=\langle\chi^{i}|\Delta\mathcal{H}|\chi^{j}\rangle.
\]
We first show that the quadratic terms vanish as follows. 
Note that
\begin{eqnarray*}
\langle\chi^{i}|\Gamma^{1}|\chi^{j}\rangle & = & \langle\chi^{i}|\Gamma^{2}\Gamma^{2}\Gamma^{1}|\chi^{j}\rangle\\
 & = & -i\langle\chi^{i}|\Gamma^{2}\left(i\Gamma^{2}\Gamma^{1}\right)|\chi^{j}\rangle\\
 & = & -i\mathbf{sgn}(\epsilon)\langle\chi^{i}|\Gamma^{2}|\chi^{j}\rangle,
\end{eqnarray*}
and
\begin{eqnarray*}
\langle\chi^{i}|\Gamma^{1}|\chi^{j}\rangle & = & \langle\chi^{i}|\Gamma^{1}\Gamma^{2}\Gamma^{2}|\chi^{j}\rangle\\
 & = & -i\langle\chi^{i}|i\Gamma^{1}\Gamma^{2}\Gamma^{2}|\chi^{j}\rangle\\
 & = & i\langle\chi^{i}|\left(i\Gamma^{2}\Gamma^{1}\right)\Gamma^{2}|\chi^{j}\rangle\\
 & = & i\mathbf{sgn}(\epsilon)\langle\chi^{i}|\Gamma^{2}|\chi^{j}\rangle.
\end{eqnarray*}
As a result, $\langle\chi^{i}|\Gamma^{1}|\chi^{j}\rangle=-\langle\chi^{i}|\Gamma^{1}|\chi^{j}\rangle$,
and thus the quadratic terms have no contribution to $\mathcal{H}_{eff}$.
For the linear terms of $\Delta\mathcal{H}$, noting that 
\[
\left[i\Gamma^{b}\Gamma^{1},\Gamma^{a}\right]=0,\qquad for\quad a\ne1\; or\; b,
\]
 all $\Gamma^{a}$s with $a\ne1\; or\; b$ can be diagonalized into
$2^{n-1}\times2^{n-1}$ blocks as 
\[
\Gamma^{a}=\begin{pmatrix}\Gamma_{+}^{a}\\
 & \Gamma_{-}^{a}
\end{pmatrix},
\]
where each block satisfies
\[
\left\{ \Gamma_{\eta}^{a},\Gamma_{\eta}^{b}\right\} =2\delta^{ab}\mathbf{1}_{2^{n-1}\times2^{n-1}},\quad a,b\ne1\; or\; b.
\]
Thus either $\left\{ \Gamma_{+}^{a}\right\} $ or $\left\{ \Gamma_{-}^{a}\right\} $
forms a $2^{n-1}\times2^{n-1}$ representation of the Clifford algebra.
Thus up to a unitary transformation, we can express the effective
theory as 
\begin{equation}
\mathcal{H}_{eff}=\frac{1}{2}\left(\mathbf{sgn}(m)+\mathbf{sgn}(\epsilon)\right)\sum_{a=2}^{d}k_{a}\Gamma_{(2n-1)}^{a+b-3},\label{eq:Effective-Weyl}
\end{equation}
subject to the condition of 
Eq.(\ref{eq:Existence-Condition}). Note that for $2\mathbf{Z}$ cases,
the first $\Gamma$ matrix in the above expression  should be $\tilde{\Gamma}$. 

It is a crucial to observate that the low-energy effective theory (\ref{eq:Effective-Weyl}),
which is obtained on the boundary of the unified model (\ref{eq:Unified-TI})
of TIs/TSCs, takes also a unified form for all types of FSs,
$i.e.$, models (\ref{eq:Weyl-type}),
(\ref{eq:First-Weyl}), (\ref{eq:Second-Weyl}), (\ref{eq:FS-2Z-nonchiral}),
(\ref{eq:Dirac-type}), (\ref{eq:FS-Dirac-Z2}), and (\ref{eq:FS-Dirac-2Z}).
Since our final aim is to prove the general index theorem, namely,
to show that the same topological information is encoded in both
bulk and boundary, our remaining task is to look into the topological
charge of the Fermi point with the obtained effective theory. To be
explicit, we shall complete two tasks: the first one is to identify the
symmetry operators of a boundary effective theory according to those
of its bulk Hamiltonian; the second one is to check whether the topological
charge of the Fermi point on the boundary matches the bulk topological
number  on both magnitude and topological type. The first one
can be done by comparing the symmetries and their signs, and thereby
it turns out that, for a given symmetry case, the corresponding
symmetry operators are just those we defined previously in the constructions
of that kind of FS with unit topological charge. This can
be clearly seen from the fact that our constructions of Fermi surfaces
and TIs/TSCs matches to our classification table of both FSs
and TIs/TSCs, where the symmetry situation of an FS of codimension
$d-1$ is the same as that of a $d$-dimensional TI/TSC. For the second
task, again from our classification table of FSs and TIs/TSCs,
we see that the topological charge of the Fermi point has the same
type as that of the corresponding bulk topological number, and furthermore
their numerical equality is obvious. Thus we finally have the following general
index theorem:
\[
\nu(i,d-1)=N(i,d),
\]
where $i$ is the index of the symmetry classes, $d-1$ is the codimension
of the Fermi point on the boundary, and $d$ is the spatial dimension, for a concerned TI/TSC.


\begin{thebibliography}{References}
\bibitem{Topo-Terms} Eligible topological terms in Ref.\cite{TIClassification-I,TIClassification-II,TIClassification-III} consisting of WZ-term and $\mathbf{Z}_2$ $\theta$-term have no adjustable parameters.
\bibitem{note} To elaborate the point more clearly, let us consider a $d$-dimensional TI/TSC with the two natural boundaries at $x_1=\pm L$. 
The boundary energy-spectrum can be obtained as $E_s(k_1, ..., k_{d})=E_{bulk}(k_1, ..., k_{d})-E_{op} (k_2, ..., k_{d})$, where $E_{op}$ is the open-boundary 
energy-spectrum that 
is solved using the open boundary condition at $x_1=\pm L$ and the periodic boundary condition for the remaining $d-1$ coordinates at $\pm L$ (if
$L \rightarrow \infty$, the result becomes exactly valid). 
 Since  $E_{bulk}(k_1, ..., k_{d})$ has a gap in the whole $\mathbf{k}$-space and the bulk gap must be closed at the two boundaries, the gapless points/regions at the boundary must be the intersecting points of the vector ${\bf{k}}_1$ and  the $E_s=0$  equal-energy surface (the Fermi energy is set to be zero here), 
and they are indeed points on the $k_{2}k_3$-plane  or some line-segments that can be continuously deformed to points for  $d=3$ (the highest dimension for realistic physical systems). As for $d>3$, from the same analysis, it can be seen that they are
at most some localized manifolds in the $\mathbf{k}$-space of the boundary, which are topologically equivalent to points.
\bibitem{Dirac-Matrix} In our notation, $\Gamma^1=\sigma_1\otimes\sigma_1$, $\Gamma^2=\sigma_2\otimes\sigma_1$, $\Gamma^3=\sigma_{3}\otimes\sigma_1$, $\Gamma^4=\mathbf{1}\otimes\sigma_2$ and $\Gamma^5=\mathbf{1}\otimes\sigma_3$, with $\sigma_i$ being Pauli matrices.
\bibitem{Note-Deformation} When we deform the bulk Hamiltonian, the boundary effective theory may also be deformed accordingly; however, the total topological charge of FSs is invariant.
\end{thebibliography}
\begin{thebibliography}{References}
\bibitem{Note-Lattice} For lattice models, the expansion near the closest points in the spectrum is just a collection of our continuous models, which characterizes the low-energy behavior.
\end{thebibliography}
\end{document}